\documentclass[a4paper,11pt]{article}
\usepackage{geometry}
\usepackage{amsmath}
\usepackage{amssymb}
\usepackage{caption}
\usepackage{graphicx}
\usepackage{latexsym}
\usepackage{times} 
\usepackage[english]{babel}
\usepackage{fancyhdr}
\usepackage{comment}
\usepackage{float}
\usepackage{xcolor}
\usepackage{soul}
\usepackage{epstopdf}
\usepackage[inkscapelatex=false]{svg}
\usepackage{enumitem}
\usepackage{subcaption}
\usepackage{hyperref}

\addto{\captionsenglish}{\renewcommand{\abstractname}{Executive Summary}}

\fancyfootoffset{1cm}
\fancypagestyle{plain}{%
\fancyhead{}
\cfoot{}
\fancyfoot{}
\rfoot{\thepage}
\lfoot{{\small EVS38 International Battery, Hybrid and Fuel Cell Electric Vehicle Symposium}}

}

\renewenvironment{abstract}
 {
  {\bfseries \large{\abstractname}}
  \par
  \vspace{10pt}
  \normalsize
 }

\title{\Large \emph{EVS38}\\ \emph{Göteborg, Sweden, June 15-18, 2025}\\ \hspace{10pt}\\ \LARGE\bf 
AI Safety Assurance in Electric Vehicles: A Case Study on AI-Driven SOC Estimation}

\author{
{\small Martin Skoglund$^1$\!, Fredrik Warg$^1$\!,  Aria Mirzai$^1$\!, Anders Thorsén$^1$\!, Karl Lundgren$^1$\!,}\\{\small{Peter Folkesson$^1$\!, Bastian Havers-Zulka$^1$}} \\
{\footnotesize $^1$\em \{martin.skoglund, fredrik.warg, aria.mirzai, anders.thorsen, karl.lundgren, peter.folkesson, bastian.havers-zulka\}@ri.se}\\ 
{\small RISE Research Institutes of Sweden, Borås, Sweden}
        }
\date{}
\begin{document}

\setlength{\parindent}{0mm}
\baselineskip 10pt

\maketitle

\vspace{12pt}

%\color{red}
%\input{WhosDoingWhat.tex}
%\color{black}

\rule{\textwidth}{1pt}
\begin{abstract}
%put abstract here
%Insert short executive summary here in 10.5pt Times New Roman. The text of the executive summary is justified with 1$1\over 2$ line spacing. The executive summary should contain between 50 and 150 words. The section shall be enclosed within two horizontal lines 1pt thick. 
Integrating Artificial Intelligence (AI) technology in electric vehicles (EV) introduces unique challenges for safety assurance, particularly within the framework of ISO 26262, which governs functional safety in the automotive domain. Traditional assessment methodologies are not geared toward evaluating AI-based functions and require evolving standards and practices. This paper explores how an independent assessment of an AI component in an EV can be achieved when combining ISO 26262 with the recently released ISO/PAS 8800, whose scope is AI safety for road vehicles. The AI-driven State of Charge (SOC) battery estimation exemplifies the process. Key features relevant to the independent assessment of this extended evaluation approach are identified.
As part of the evaluation, robustness testing of the AI component is conducted using fault injection experiments, wherein perturbed sensor inputs are systematically introduced to assess the component's resilience to input variance.
\end{abstract}

\rule{\textwidth}{1pt}
\vspace{10pt}

\section{Introduction}

ISO 26262 \cite{iso26262} is the international standard for functional safety (FuSA) in road vehicles, focusing on reducing risks from system malfunctions through structured processes and safety mechanisms. For automated driving/advanced driver assistance systems (AD/ADAS) there is also the complementary ISO 21448 safety of the intended functionality (SOTIF) standard \cite{iso21448}. However, these lacks specific guidance for addressing AI safety, particularly in managing AI's black-box nature and data dependency. The black-box nature of AI limits transparency and traceability in decision-making, while its data dependency means that the quality and representativeness of training data significantly influence system behaviour. These characteristics create gaps in traditional safety assessments, requiring new approaches to evaluate and ensure the safety of AI-driven systems.

This paper explores expanding FuSA assessment evaluations using the new ISO/PAS 8800 \cite{ISOPAS8800}, which discusses safety and artificial intelligence in road vehicles. This extension addresses gaps by identifying and mitigating systematic faults and insufficiencies in AI systems to ensure safety during development and beyond. To this end, one may conduct independent assessments on an assurance case on a heterogeneous set of evidence from various test environments that align with multi-pillar methodologies such as those described in NATM (new assessment/test method for automated driving) \cite{natm} %and the large-scale EU project SUNRISE \cite{sunrise}
, which are relevant for AD/ADAS systems.

The main challenges in assessing AI systems within a safety assurance framework include evaluating performance robustness and identifying performance insufficiencies. Performance robustness refers to the system's ability to maintain safe operation under expected variations and disturbances within the operating conditions. Performance insufficiency occurs when AI systems fail in unanticipated scenarios due to technical limitations. Addressing these issues is critical in FuSA and SOTIF to ensure that systems meet essential performance standards, even in dynamic or unpredictable environments. Systematic faults in AI design further complicate safety assessments. AI systems trained on operating data are prone to subtle design flaws that may introduce hidden hazards, which may only become evident under specific conditions. Clear guidance is therefore needed to manage these risks, ensuring that AI systems remain robust throughout the design and deployment phases.

This work extends conventional FuSA assessment for AI-based state of charge (SOC) systems to address AI-specific challenges, including training data quality and validity. It identifies safety cages as key architectural elements aligned with ISO 26262 and proposes them as a natural interface for integrating ISO/PAS 8800 to address core AI-related measures. Robustness evaluation is conducted using fault injection experiments, where systematically perturbed sensor inputs are applied to investigate an AI component’s failure characteristics. This can provide evidence of the AI components' behavior under abnormal and unforeseen conditions, thereby addressing some inherent uncertainty in assessing AI-based components.

%Remved picute due to space issues, need at least one picture?
%\begin{figure}[htbp!]
%\begin{centering}
%\includegraphics [width=0.99\textwidth]{fig/Concerns_old.pdf}
%\caption{\hl{ UPDTE this image with an image of an assurance case in GSN?}}
%\label{concerns}
%\end{centering}
%\end{figure}

%\section {The evaluation process for AI within the safety engineering context}

%ISO/PAS 8800 uses a simplified scheme for error classification compared with the random hardware faults as described in ISOP 26262-5. The scheme is illustrated in  Figure \ref{fig:errorclassification} and is based on the potential to lead to an undesired safety-related behavior at vehicle level.

%\begin{figure}[htbp!]
%	\centering
%	\includegraphics[width=10cm]{fig/ErrorClassification.drawio}
%	\caption{AI error classification scheme from ISO/PAS 8800 \cite{ISOPAS8800}. Red indicates contributing AI error and green non-contributing error.}
%	\label{fig:errorclassification}
%\end{figure}

\section{Integrating AI in Safety Assurance}

\begin{figure}[htbp!]
	\centering
	\includegraphics[width=0.7\linewidth]{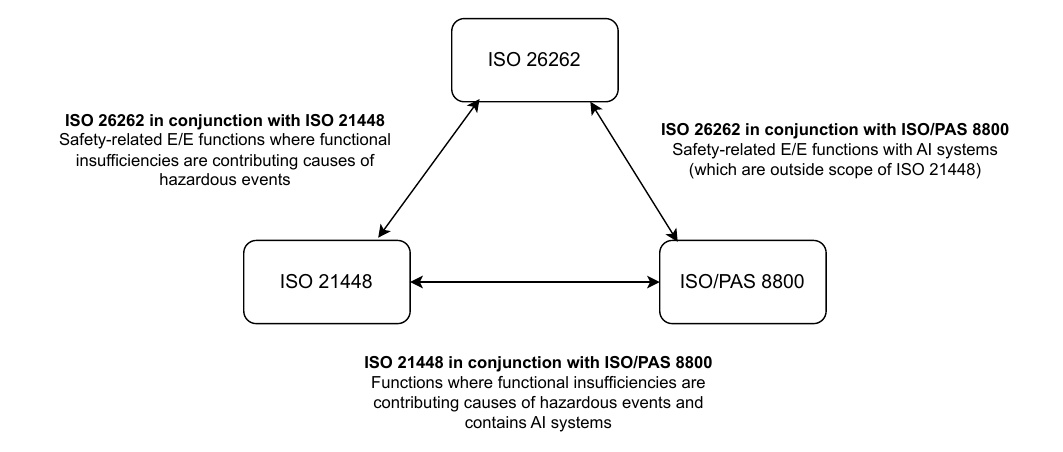}
	\caption{Relationship between road vehicle safety standards ISO 26262, ISO 21448, and ISO/PAS 8800 for use when developing functions with AI components.}
	\label{fig:standardsrelation}
\end{figure}

Several safety-related standards for electric/electronic (E/E) systems in road vehicles may apply to the context of AI systems. The primary safety standard, ISO 26262, covers functional safety, including systematic and random hardware faults for all E/E systems in road vehicles. For systems with complex sensors, such as cameras, lidar, or radar, which is especially relevant for ADAS and AD functions, ISO 21448 \cite{iso21448} additionally covers functional insufficiencies. These include performance limitations in technical abilities (e.g., sensor performance) or insufficiencies in the specification, when either of these insufficiencies can lead to hazardous behaviour under some relevant conditions (triggering conditions). The safety of AI systems is covered in ISO/PAS 8800, which is developed to be used in conjunction with the two aforementioned standards. Figure \ref{fig:standardsrelation} illustrates how these standards are interrelated, i.e., depending on the function under development, using two or all three may be necessary. This paper and the SOC case study focus on ISO 26262 and ISO/PAS 8800, which are sufficient for a system without complex sensors. As mentioned above, a similar integration must include ISO 21448 for systems with complex sensors.

\begin{figure}[h]
	\centerline{\includegraphics[width=7.5cm]{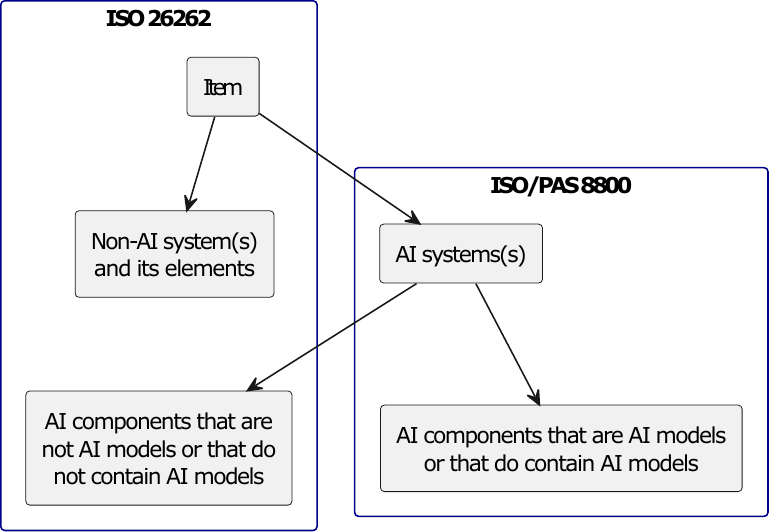}}
	\caption{Illustration of applicability of ISO 26262 respectively ISO/PAS 8800 (based on  \cite{ISOPAS8800}).}
	\label{fig:26252_8800}
\end{figure}

A first step to integrate AI in safety assurance is to define for which parts ISO 26262 is still applicable and when it needs to be tailored to use ISO/PAS 8800 \cite{ISOPAS8800}. The latter defines an \emph{AI system} as a top-level abstraction for AI-based functionality, an element containing one or more \emph{AI components}. An AI component may be a pre- or post-processing component or an \emph{AI model}, where the latter is a construct making inferences based on some input, e.g., a trained deep neural network with its weights and hyper-parameters. As illustrated in Figure \ref{fig:26252_8800}, the overall AI system and AI models fall within the scope of ISO/PAS 8800. In contrast, the full \textit{item}\footnote{Item is the ISO 26262 term for a function at vehicle level that falls within its scope, i.e., a function containing E/E elements.} and all non-AI system elements as well as AI components that are not AI models fall within the scope of ISO 26262. For instance, a pre-processing component that is not AI-based will be assessed under ISO 26262, even if it is also part of an AI system.

\begin{figure}[h]
    \centering
    \includegraphics[width=0.95\linewidth, trim = 0cm 0cm 0cm 0cm, clip]{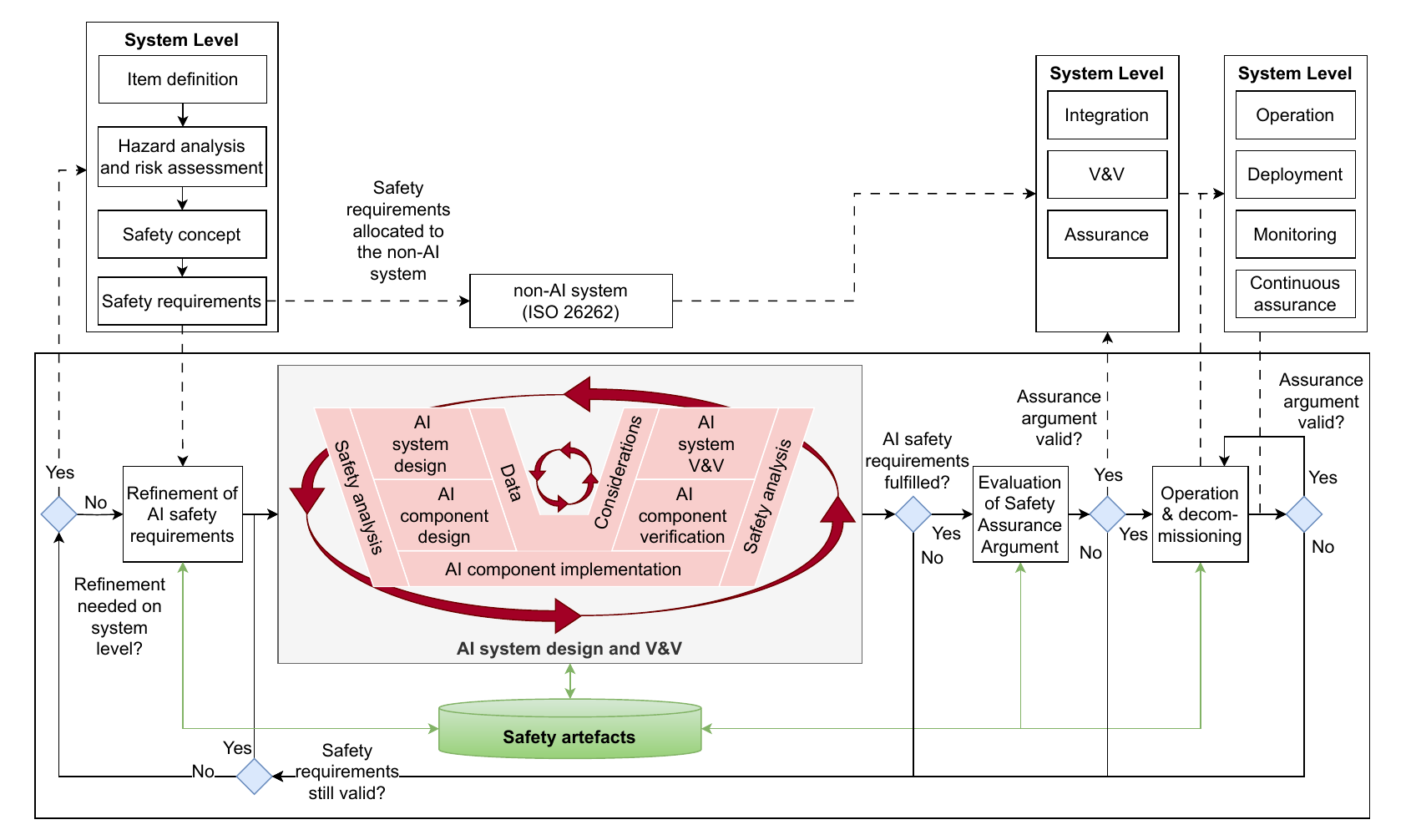}
    \caption{View of lifecycle for vehicle functions containing AI systems (based on \cite{iso26262,ISOPAS8800,henriksson_out--distribution_2023}).}
    \label{fig:ml-lifecycle}
\end{figure}

Based on this division, ISO/PAS 8800 proposes an AI reference lifecycle shown in Figure \ref{fig:ml-lifecycle}. It is based on the ISO 26262 development cycle, where safety requirements from the item under development are decomposed and allocated to the %AI
system during the system development phase. ISO/PAS 8800 tailors the ISO 26262 development cycle by adding that AI-related safety requirements are allocated to the AI system as illustrated on the left side of Figure \ref{fig:ml-lifecycle}.% including the AI components. 
 These requirements may also need to be adjusted during development and continuous assurance activities during operation %(right side of the V-model)
in cases where the AI system cannot meet its safety requirements and related properties. Some of the reasons for this could be (i) difficulties in finding suitable training and testing data, (ii) limitations in the ability to generalize to new operating conditions, or (iii) insufficient evidence to demonstrate confidence in compliance with safety standards. These challenges require iterative feedback between the AI system and the encompassing system’s safety concept and requirements.

%\hl{Anders har arbetat med texten hit - men inte nödvändigtvis färdig med det :-|}

\subsection{System-level design and verification}
In the development lifecycle, ISO 26262 \cite{iso26262} would verify system-level artifacts such as technical safety concepts, safety requirements, architectural design, and architectural measures.

System-level design can also be adapted to increase the feasibility of fulfilling requirements allocated to AI systems. This can be achieved by, e.g., restricting the allowed operating conditions, incorporating diverse processing algorithms or sensing modalities, or implementing redundancy.

%When conducting verification and validation of safety requirements, the confidence in the assessment results will be high when all the three road vehicle standards of interest (ISO 26262 \cite{iso26262}, ISO 21448 \cite{iso21448}, and ISO/PAS 8800 \cite{ISOPAS8800}) are followed. To this end, ISO 26262 \cite{iso26262} and ISO 21448 \cite{iso21448} could be used to verify technical safety concepts and safety requirements of AI systems. In contrast, ISO 26262 \cite{iso26262} could also be used to verify architectural design, models or architectural measures. ISO/PAS 8800 \cite{ISOPAS8800} would then need to be followed to verify data's correctness, consistency, and completeness when performing systematic tests for evaluating work products and elements within a test environment at different levels of system integration.

%AI systems would also need to undergo a thorough verification and validation (V\&V) process to ensure the system's safety under test. In this context, verification assures that the specified requirements have been fulfilled, while validation assures that the requirements for a specific intended use have been fulfilled. According to ISO/PAS 8800 \cite{ISOPAS8800}, methods used to verify AI system safety may also be used for AI system safety validation.

\subsection{AI system verification} \label{8800techniques}
A key challenge in verification for most types of AI systems, including the most common type of machine learning (ML), is that the AI model is largely opaque, i.e., it is not possible to inspect the model or use formal methods to determine if it fulfills the safety requirements. An inherent limitation of ML models, in particular, stems from their reliance on training data, where data quality issues or training method issues can result in unpredictable behaviour, especially in edge cases or rare conditions that deviate from the training data distribution. Addressing this issue necessitates appropriate verification methods but typically also robust fallback safety mechanisms capable of mitigating the potential consequences of such unexpected behaviour. 

%One would need to address several challenges to verify and validate an AI system, some of which are discussed here. Given that the inputs to AI systems are from several different sources and rely on large datasets, covering and exploring all combinations of values for these inputs is challenging, which could cause issues centred around the completeness of verification results.
%Minor changes in these inputs as well as changes introduced as a result of re-training can cause unpredictable impact (in some cases even errors) on the verification results.
%On the other hand, some of the inputs used by AI systems (e.g., traffic signs and driver behavior) are even difficult to represent mathematically. Apart from the diversity and size of inputs which causes scalability issues during the verification process, the large number of parameters used by neural networks (NN) that are used e.g., in perception-equipped systems, could cause scalability issues. Examples include weight parameters which typically have a significant influence on the results generated by these networks. AI systems, in general, could also get stuck in local optimums due to inadequate examples while training instead of producing globally optimum results.

Thus, choosing appropriate combinations of test methods to address these challenges becomes important. Preferably, traditional safety-related test methods should be complemented with test methods specifically targeting AI systems. Some examples of such test methods are:

\begin{itemize} 
    \item Gradient-based search methods using analysis of the AI model to guide the generation of test cases.
    \item Statistical testing, i.e., evaluation of performance metrics measurable on the model under test, e.g., precision and recall, with the desired confidence interval.
    \item Test cases designed with expert domain knowledge or based on reviews of the used model and data set.
    \item Robustness testing, e.g., applying noise patterns or other disturbances in the input to measure the model's resilience to input variance.
    \item Use of explainability techniques for making the model's decisions semi-transparent. It can be used to understand how a model works and identify potential weaknesses or biases.
    \item Cross-validation, i.e., dividing the available dataset into several training/verification tuples to test if the model is robust for different training sets.
    \item Sampling-based methods that can guide testing to areas in the input space with higher error probability.
\end{itemize}

There may be challenges in the physical testing of AI systems operating in complex environments, particularly for achieving sufficient coverage of edge cases. In those cases, virtual testing can be used to complement physical testing. Virtual test platforms may also facilitate the generation of synthetic datasets to address the challenges of achieving adequate distribution and coverage of the inputs to AI systems. However, this requires the entire generation workflow to be validated and correlation with real data to be made. 

%A precise way of testing the accuracy of the synthetic datasets is through Hardware-in-the-Loop (HiL) testing. By reproducing the exact same scenes recorded from the real world in the virtual world injected into real sensors of a HiL rig, a comparison between the real and virtual dataset can be made.

\subsection{AI system validation}

In addition to field testing, virtual testing techniques used for verification may also be used for validation once integrated into the encompassing system, where they can be used to explore relevant scenarios systematically and identify corner cases or abnormal situations. Methods for detecting out-of-distribution data, i.e., input data that is not similar to the training data, can aid the evaluation \cite{henriksson_out--distribution_2023}.

%to determine whether an input can be included in the training dataset by considering its differences with the existing training data.

\subsection{AI system safety analysis}

For the safety analysis of AI systems, the aim is to provide confidence that the risk of violating the AI safety requirements at the AI system level due to AI errors is sufficiently low. 
%ISO/PAS 8800 states that safety analysis of AI systems complements safety analysis in accordance with ISO 26262 and ISO 21448, and is performed during the AI system design and V\&V as well as AI component design, implementation, and verification. Thus, the choice of analysis techniques should focus on techniques that identify the safety-related AI errors of the AI models in AI systems and their components, while in the case of errors affect components of non-AI models, ISO 26262 and ISO 21448 applies.
The safety analysis techniques should adequately identify hazards and their potential causes. Some off-the-shelf techniques may be reused or enhanced to analyse AI systems. Examples of such techniques include fault-tree analysis (FTA) \cite{ali_analyzing_2020}, failure mode and effects analysis (FMEA) \cite{salay_safety_2019}, and hazard and operability analysis (HAZOP) \cite{qi_hierarchical_2022}. While these techniques analyse systems with certain underlying assumptions, other state-of-the-art techniques have been introduced with stronger assumptions to model AI systems \cite{adee_systematic_2021,adee_discovery_2021,berk_assessing_2020}.

If AI errors are identified as a result of testing, analyses are performed to evaluate their impact. Typically, the analysis activities include risk evaluation, root-cause analysis, and risk mitigation. Risk evaluation involves assessing the risk of a failed test to estimate the impact on safety. Root-cause analysis involves identifying the underlying issues for the AI errors, which may be related to AI safety requirements, datasets, or AI model design. After the risk evaluation and root-cause analysis have been performed, the risks are mitigated through the definition of prevention, detection, and control measures for the identified root causes. Thus, depending on the root cause, the mitigations may involve changes to the AI safety requirements, AI model, dataset, or the AI development process.

\section{Case study: AI-based SOC estimator for EV battery}
% \begin{figure}[htbp!]
% \begin{centering}
% \begin{minipage}{.65\linewidth}
% \includegraphics[width=\linewidth]{fig/Traditional_nomeasures.eps}

% \caption{State of charge component with traditional estimation.}\label{figsoc_trad}
% \end{minipage}
% \begin{minipage}{.34\linewidth}
% \includegraphics[width=\linewidth]{fig/SocAI_nomeasures.eps}

% \caption{State of charge AI-based estimation.}\label{figsoc_est2}
% \end{minipage}
% \end{centering}
% \end{figure}

\begin{figure}[htbp!]
     \centering
     \begin{subfigure}[t]{0.61\textwidth}
         \centering
         \includegraphics[width=\linewidth]{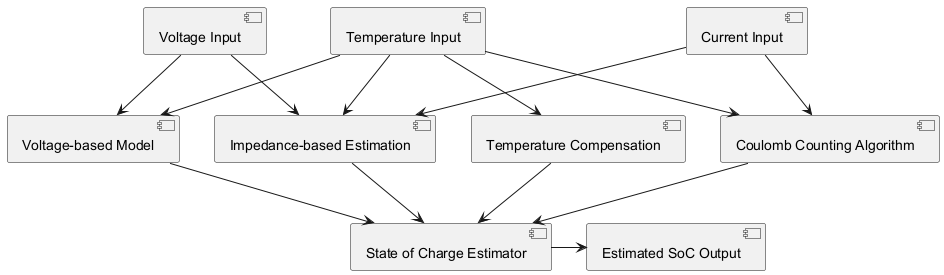}
         % \caption{State of charge component with traditional estimation.}
         \caption{}
         \label{figsoc_trad}
     \end{subfigure}%
     \begin{subfigure}[t]{0.38\textwidth}
         \centering
         \includegraphics[width=\linewidth]{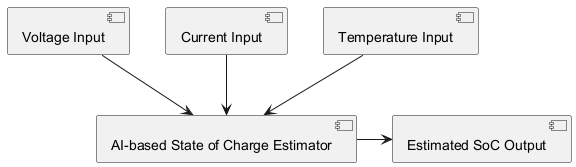}
         % \caption{State of charge AI-based estimation.}
         \caption{}
         \label{figsoc_est2}
     \end{subfigure}
     \caption{SOC estimation methods: Traditional estimation (\ref{figsoc_trad}) and AI-based estimation (\ref{figsoc_est2}).}
\end{figure}

The SOC (state of charge) measures the remaining charge in a battery, typically expressed as a percentage of its total capacity. In EVs, SOC provides critical information for range estimation and safety functions, such as preventing overcharging and deep discharging in battery management systems (BMS). Failure to accurately measure SOC can result in overcharging, potentially causing excessive heat generation, electrolyte decomposition, and, in extreme cases, thermal runaway. Estimating SOC in batteries is challenging due to their nonlinear behavior and dependency on operating conditions such as temperature, aging, and discharge rates \cite{demirci_review_2024}. Figure \ref{figsoc_trad} shows a traditional SOC estimation method that relies on coulomb counting, which measures the charge entering and leaving the battery; open-circuit voltage analysis, which maps the battery's voltage at rest to its SOC; and physics-based electrochemical models. Recently, however, AI-based SOC methods have gained traction due to their ability to model batteries' complex and nonlinear behaviour; such an estimator is illustrated in Figure \ref{figsoc_est2}.

\subsection{AI-based SOC estimator with monitor}
\label{ai_monitor}
% \begin{figure}[htbp!]
% \begin{centering}
% \begin{minipage}{.45\linewidth}
%   \includegraphics[width=\linewidth]{fig/Monitor_actuator.eps}
%    \caption{Actuator/Monitor Architecture safety pattern.}
%    \label{figpattern1}
% \end{minipage}
% \begin{minipage}{.55\linewidth}
% \includegraphics[width=\linewidth]{fig/SocAI_withmeasure.eps}
%    \caption{Actuator/Monitor pattern applied to AI SOC estimator in Fig. \ref{figsoc_est2}.}
%    \label{figpattern2}
% \end{minipage}
% \end{centering}
% \end{figure}
\begin{figure}[htbp!]
     \centering
     \begin{subfigure}[t]{0.39\textwidth}
         \centering
         \includegraphics[width=\linewidth]{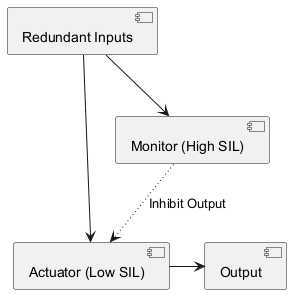}
   % \caption{Actuator/Monitor Architecture safety pattern.}
   \caption{}
   \label{figpattern1}
     \end{subfigure}%
     \begin{subfigure}[t]{0.60\textwidth}
         \centering
         \includegraphics[width=\linewidth]{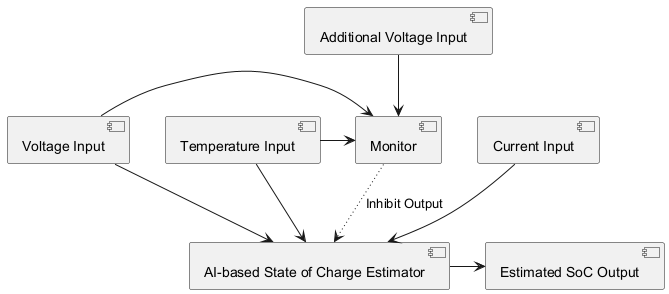}
   % \caption{Actuator/Monitor pattern applied to AI SOC estimator in Fig. \ref{figsoc_est2}.}
   \caption{}
   \label{figpattern2}
     \end{subfigure}
     \caption{Actuator/Monitor Architecture safety patterns: General (\ref{figpattern1}, where SIL = safety integrity level) and applied to AI SOC estimation  (\ref{figpattern2}) from Figure \ref{figsoc_est2}.}
\end{figure}
A well-known safety pattern is to encapsulate a more complex function, where ensuring safety is difficult, by adding a simpler monitor to restrict the output of the complex function to safe ranges. The complex component will, in normal operation, provide better performance. In comparison, the monitor intervenes when the complex component fails but provides lower performance or keeps the system safe. Figure \ref{figpattern1} illustrates this pattern. Input signals go to the complex component (actuator) and the monitor. There are some variations of the pattern, but in this example, the monitor can inhibit the actuator's output if the output calculated by the actuator component is outside the safe range. The advantage of the pattern is that the safety requirements can be allocated to simpler components that are feasible to develop with a high safety integrity level (SIL). However, it is important to realize that the monitor must be able to intervene in all failure modes of the complex component. If not, there will be a shared responsibility where the complex component will still be assigned some safety requirements.

In AI, this pattern is sometimes called safety envelope or safety cage \cite{neto2022} and means there is a non-AI component monitoring the AI component. Figure \ref{figpattern2} shows our case study SOC estimator using this pattern to handle some of the fault modes; a monitor uses voltage and temperature to be able to inhibit the SOC output such that thermal runway, the most critical consequence of erroneous SOC estimation, can be avoided.

%Building on the findings by Neto et al. \cite{neto2022}, we introduce the safety cage as an architectural element that encapsulates redundancy, isolation, monitoring, fault detection, and fallback mechanisms to ensure critical functions remain safe during faults or AI deviations, aligning with ISO 26262 and supporting AI-focused standards like ISO/PAS 8800. The objective is to evaluate whether the safety cage-relevant measures in Table~\ref{tab:safety_mechanisms_with_tests} fulfil their intended purpose under ISO 26262, leveraging traditional, AI-independent design. Conversely, the testing and assessment of non-safety cage measures will likely need to be replaced, complemented, or aligned with the measures and test methods in the recently published ISO/PAS 8800.

\subsection{AI-Specific Safety Assessment Features for SOC} \label{ai_arch}
The SOC estimator functions as a sensor and is subject to the same typical failure modes as conventional sensors. A detailed analysis is necessary to ensure its safety and proper functionality. As outlined in ISO 26262-5 Table D.9, the typical failure modes for sensors include \textbf{out-of-range signals, offset errors, signals remaining stuck within a valid range, and oscillatory behaviour}.

In addition to these standard failure modes, AI-based SOC estimation systems introduce at least two additional assessment considerations due to their potential contribution to the abovementioned failure modes.

\textbf{Training Data Quality and Relevance: }The performance of AI systems heavily depends on the quality of their training data. The data must be relevant, sufficiently representative, and as complete and error-free as possible. Inadequate or biased training data can lead to poor generalisation and incorrect SOC estimates.

\textbf{Validity of Training Data Over Time: }Training data is inherently limited to the conditions under which it was collected and cannot guarantee universal validity in future operational scenarios. To mitigate this, in-service monitoring must continuously verify that the assumptions and assertions made during training remain valid in the operational environment.

Given the fault modes and unique characteristics of AI systems, ensuring the safety of an AI-based SOC estimation system requires a careful assessment to confirm the existence of appropriate safety measures and evidence of their effectiveness, i.e., test results, that address both traditional and AI-specific challenges. The selection of test methods (Table~\ref{tab:safety_mechanisms_with_tests}) for each safety mechanism is based on their specific objectives in ensuring system safety. Methods for obtaining evidence of the correct implementation of functional and technical safety requirements at the system level are selected from ISO 26262-4, Table 9. Similarly, methods for validating correct functional performance, accuracy, failure mode coverage, and the timing of safety mechanisms at the system level are chosen from ISO 26262-4, Table 10.

\begin{figure}[h]
    \centering
    \includegraphics[width=1.0\linewidth]{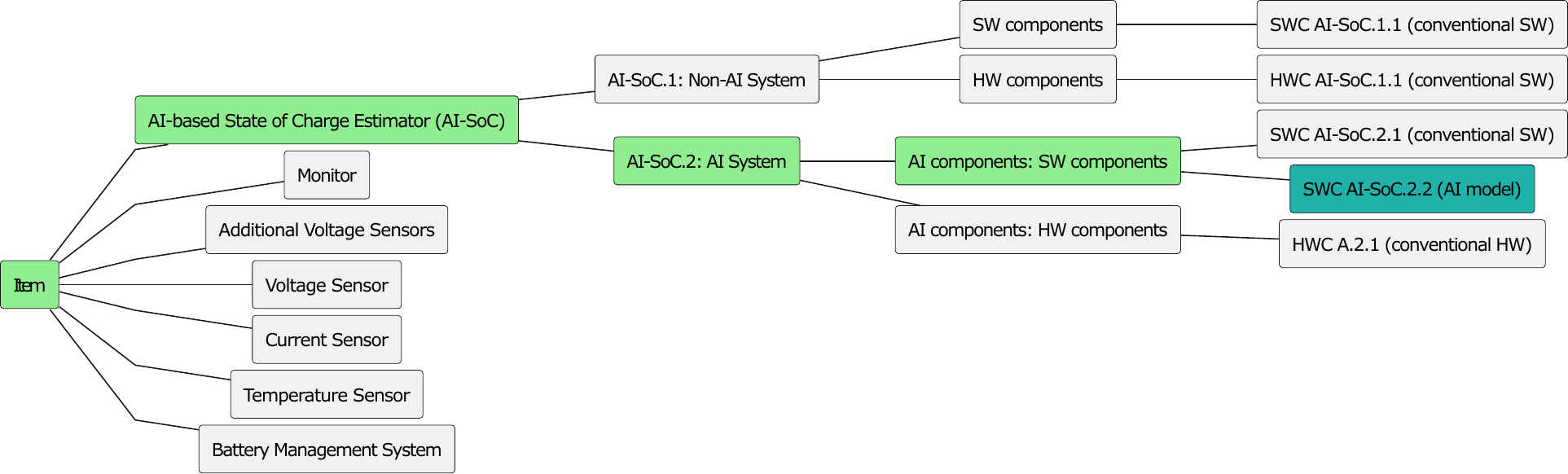}
    \caption{Hierarchical decomposition of the AI SOC estimator in Figure \ref{figpattern2}. The green colour indicates elements subject to the ISO/PAS 8800 tailored process.}
    \label{fig:decomposition}
\end{figure}

\begin{table}[h!]
\small
\caption{Safety Mechanisms, traditional Test Methods, and Assessment Aims adapted from ISO 26262-5 Table D.9 — Sensors}
\centering
\begin{tabular}{|p{0.15\textwidth}|p{0.30\textwidth}|p{0.50\textwidth}|}
\hline
\textbf{Measure Name} & \textbf{Test Methods} & \textbf{Assessment Aim} \\ \hline

%Failure detection by online monitoring&  Fault Injection Test, Performance Test & Assess ability to detect deviations in system behaviour during normal operation and handle AI model anomalies, e.g. signals stuck within a valid range and oscillatory behaviour.  \\ \hline

%Test pattern &   Requirement-Based Tests, Back-to-Back Tests & Assess the ability to detect static failures and deviations by comparing AI outputs to expected patterns, e.g. signals stuck within a valid range. \\ \hline

%Sensor valid range &  Fault Injection Test, Requirement-Based Test & Assess ability to detect out-of-range inputs and validate SOC estimation for invalid data, e.g. out-of-range signals. \\ \hline

Input comparison/voting & Fault Injection Test, Error Guessing Test & Assess ability to detect discrepancies across redundant inputs or models, e.g., offset errors and signals stuck within a valid range. \\ \hline

Sensor correlation & Performance Test, Fault Injection Test & Assess ability to detect inconsistencies between sensors and mitigate sensor drifts in SOC estimation, e.g., offset errors and oscillatory behaviour. \\ \hline

Sensor rationality checks & Error Guessing Tests derived from Field Experience & Assesses the ability to detect implausible outputs and maintain SOC plausibility using diverse inputs, e.g., offset errors and out-of-range signals. \\ \hline

\end{tabular}
\label{tab:safety_mechanisms_with_tests}
\end{table}
When mapping safety measures for an SOC sensor implemented according to the architectural pattern depicted in Fig \ref{figpattern1}, some responsibilities fall to the monitor component and can, therefore, be addressed using traditional functional safety measures and test techniques listed in Table \ref{tab:safety_mechanisms_with_tests}. However, as listed below, certain sensor-related AI-concerns (AIC) pertain specifically to the AI-based SOC estimation. Thus, they require the integration of an ISO/PAS 8800 tailored process, depicted in Figure \ref{fig:decomposition}, and test techniques selected from Section \ref{8800techniques} to provide evidence of safety objectives fulfillment.

\begin{enumerate} [label=\textbf{AIC\arabic*}]
\item Detecting failures through \textit{online} monitoring, emphasizing evaluating the system's ability to identify deviations in behaviour during normal operation and to manage anomalies in the AI model. This includes detecting conditions such as \textbf{signals remaining stuck within a valid range} and \textbf{oscillatory behaviour}. \label{AIC1}

\item Assessing the AI model’s capability to detect \textit{static} failures and deviations by employing test patterns that compare the AI's output against expected behaviours. Relevant fault modes include \textbf{signals remaining stuck within a valid range} and \textbf{offset errors}. \label{AIC2}

\item Validating sensor input ranges by verifying the system’s ability to detect and appropriately respond to \textbf{out-of-range signals}, ensuring that SOC estimation remains reliable despite invalid sensor data. \label{AIC3}
\end{enumerate}

The list of concerns was compiled through a systematic assignment of each measure specified in ISO 26262-5 Table D.9 to either the monitor or the actuator based on their functional relevance. It subsequently identifies the measures and fault modes that the AI-based component of the sensor should address. In this context, particular attention is given to fault modes that may result in underestimating the state of charge.  % i.e. \textbf{signals remaining stuck within a valid range} and \textbf{offset errors}, 
This condition is closely associated with the most severe hazard, battery outgassing in the passenger compartment.

% \begin{figure}[htbp!]
% \begin{centering}
% \includegraphics [width=0.99\textwidth]{fig/out/SocAI_nomeasures.png}
% \caption{State of Charge component with AI Estimator.}
% \label{pattern}
% \end{centering}
% \end{figure}

% \begin{figure}[htbp!]
% \begin{centering}
% \includegraphics [width=0.99\textwidth]{fig/out/SocAI_withmeasure.png}
% \caption{State of Charge component with Monitor/Actuator pattern.}
% \label{pattern}
% \end{centering}
% \end{figure}

\subsection{Experiments}
Robustness testing, described in Section \ref{8800techniques}, such as applying noise patterns or other disturbances to the inputs to evaluate the model's resilience to input variance as described in \ref{AIC2}, constitutes an appropriate approach to provide evidence for the mitigation of failures related to the underestimation hazard. An example of such testing is investigated in the following Sections. 

\subsubsection{Experimental method}
\textit{Fault injection test} is a common method for evaluating safety measures, which involves accelerating the occurrences of faults for evaluating the dependability and cybersecurity properties of systems \cite{avizienis_basic_2004}. Fault injection is performed by inserting artificial faults or errors into the system, often using simple fault models such as \textit{stuck-at-0} and \textit{stuck-at-1} for permanent faults, which set the logic values to 0 or 1 respectively, and \textit{bit-flips} which are typically used for transient faults, where the logic values are inverted. The injected faults typically correspond to operational faults such as shortcuts, breaks, electromagnetic disturbances, radiation particle strikes, etc. Still, they may also correspond to development faults, such as programmers’ mistakes or flaws in semiconductor devices. In the case of evaluating cybersecurity properties of systems, fault injection may also be referred to as attack injection \cite{sangchoolie2018}.

Fault injection may be performed at many different abstraction levels and design stages, depending on the availability of system models or physical prototypes. Common fault injection techniques include simulation-based and physical techniques. Simulation-based techniques are typically used at early development stages where faults may be injected into hardware-, software- or system models, i.e., model-level fault injection. Physical fault injection is used to inject faults at the hardware level at later development stages when the actual physical system or prototype is available. Common physical techniques include pin-level- and radiation-based fault injection and other methods, e.g., using debug/test logic, EMI, or power supply disturbances. Additional software for injecting faults is typically referred to as software-implemented fault injection (SWIFI). SWIFI is an attractive technique commonly used for its flexibility and cost-effectiveness compared to other techniques. Two main approaches are used: \textit{Runtime injection}, which injects faults during system operation, and \textit{pre-runtime injection}, which injects faults before system operation. 

The stuck-at fault model is commonly used in fault injection experiments and is often applied on a single logic value (or bit), which is permanently set to 0 (stuck-at-0) or 1 (stuck-at-1) in the target system for each experiment. Apart from being a common fault model used for emulating the effects of permanent hardware faults, stuck-at faults are also part of the failure modes for sensors outlined in ISO 26262-5 Table D.9. A survey of papers from five major conferences on dependability (DSN, ISSRE, SafeComp, PRDC and EDCC) published during the last 6 years (2019-2024)\footnote{The survey in question is not yet published.} reveals that SWIFI is commonly used for evaluating AI-based systems and that that the stuck-at fault model is often used, e.g., \cite{qutub_hardware_2022,beyer_online_2023,ruiz_zero-space_2024}.

%\\cite{chen_tensorfi_2020}\cite{campos_online_2023}\cite{agarwal_resilience_2023}\cite{ding_-line_2019}\cite{qutub_hardware_2022}\cite{schmedding_strategic_2024}\cite{laskar_characterizing_2022}\cite{ruiz_zero-space_2024}.

\paragraph{Conducted experiment}
Following the above paragraph, we chose to perform our fault injection experiments on the AI-driven SOC estimation system using pre-runtime SWIFI with the stuck-at-fault model applied to multivariate input signals of an AI SOC estimation system. To detect aberrations, we compare predictions for the SOC from original data with those from data corrupted via stuck-at fault injection using absolute deviation and Root Mean Squared Error (RMSE).

\subsubsection{System under test} 

%describe practical experiments with the component architecture from the suggested Section in \ref{ai_arch} and the data set of... \\

% Conduct and describe practical experiments. Input: tied to prototype n lifecycle. Output: Experiment findings evaluating AI SOC estimator.

The experiments investigate how pre-runtime injection of stuck-at faults into the test data affects the model's SOC estimations. The injected bits of the test data are part of \texttt{Float64} values representing "Voltage," "Current," or "Temperature" inputs, which the model receives at each time instance (referred to as "steps").

The specific model used for the experiments in this paper is a Recurrent Neural Network (RNN) with Long Short-Term Memory (LSTM) cells, as presented in \cite{10.1145/3462203.3475878}. This model generates a SOC percentage prediction by processing the information from a fixed number (\textit{N}) of preceding steps. Consequently, one prediction is generated every \textit{N}th step, independent of other predictions made. Model parameter values shown to provide good performance with uncorrupted data were selected for the experimental setup. The model performs well with room-temperature data (25°C) and when trained to process 300 steps of preceding information for each SOC prediction. A pre-trained model with these specifications, along with a Python implementation for SOC estimation, is provided by the authors of \cite{10.1145/3462203.3475878} and utilised in this paper. The implementation normalises input (Voltage, Current, Temperature) values from the dataset to be within the range of zero to one before the model receives them. Said model has been trained and tested using the "LG 18650HG2 Li-ion Battery" dataset, available at \cite{Mendeley}. Although the training data only consists of six mixed discharge cycles, we believe that the results obtained with the available dataset are illustrative for cycles involving charging as well. In any event, incorrect SOC estimations may have potential safety implications, e.g., if the estimated SOC is too low at the end of a discharge cycle when the charge cycle begins.

Single stuck-at-0 or stuck-at-1 faults are injected at the start of the entire discharge cycle for each experiment. The faults are systematically injected into bits 3 to 64 for each 64-bit floating point value of the inputs (Voltage, Current, Temperature). The initial two bits are exempted, as the resulting value of the injected float may be large enough to trigger exceptions in the programming code rather than affecting the model's SOC output.

\subsubsection{Results}

\begin{figure}[htbp!]
     \centering
     \begin{subfigure}[t]{0.5\textwidth}
         \centering
         \includegraphics[width=\linewidth]{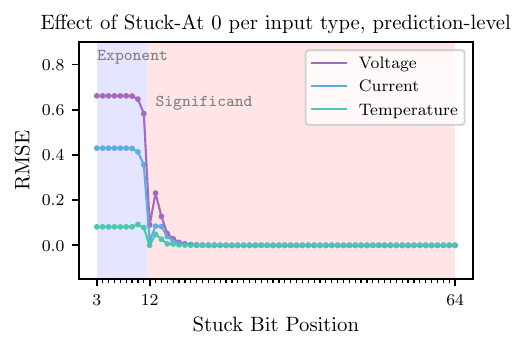}
     \caption{}
     \label{fig:rmse0}
     \end{subfigure}%
     \begin{subfigure}[t]{0.5\textwidth}
         \centering
         \includegraphics[width=\linewidth]{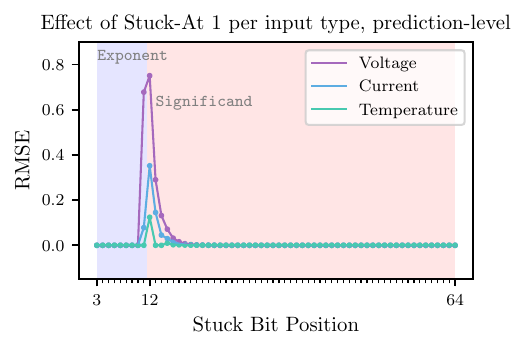}
     \caption{}
     \label{fig:rmse1}
     \end{subfigure}%
     \caption{RMSE between SOC predictions from original and corrupted data for stuck-at faults injected into bits 3-64 of input type sensor, which are in turn a series of \texttt{Float64}. \ref{fig:rmse0}: stuck-at 0; \ref{fig:rmse1}: stuck-at 1.}
     \label{fig:rmse}
\end{figure}

Figures \ref{fig:rmse0} (stuck-at-0) and \ref{fig:rmse1} (stuck-at-1) displays the results for each input type, where the experiment consisted of injecting a stuck-at fault into a single bit in the very beginning of the discharge cycle, and comparing the resulting SOC predictions with those from the uncorrupted data. Each point in the plot represents the RMSE value for the whole discharge cycle. Whether the flipped bit was in the \textit{exponent} or \textit{significand} of the \texttt{Float64} is highlighted in this plot, since the former has a greater impact on the floating point's value\footnote{The \texttt{Float64} standard is specified in IEEE 754: Bit 1 is the sign, bits 2-12 are the exponent, and the remaining bits compose the significand or mantissa}.
In this experiment, it appears that the sensors differ in how much they affect the model's predictions, with fault injections in "Voltage" causing the largest errors, followed by "Current" and finally "Temperature". Furthermore, bits in the significand have a starkly decreasing (with bit index) influence on the results.
As the data was normalised in \cite{10.1145/3462203.3475878} to the range $[0;1]$, bits 3 to 10 all have the value '1' for all sensors and at every step. Thus, stuck-at 1 has no effect for these bits, and effects only become visible for bits 11 and higher.
For stuck-at 0, we see the reverse effect, as now making the first bits in the exponent stuck at '0' leads to significant RSME of the predictions.

\begin{figure}[htbp!]
     \centering
     \begin{subfigure}[t]{0.5\textwidth}
         \centering
         \includegraphics[width=\linewidth]{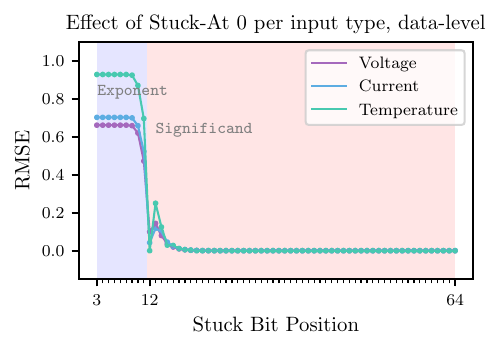}
     \caption{}
     \label{fig:data0}
     \end{subfigure}%
     \begin{subfigure}[t]{0.5\textwidth}
         \centering
         \includegraphics[width=\linewidth]{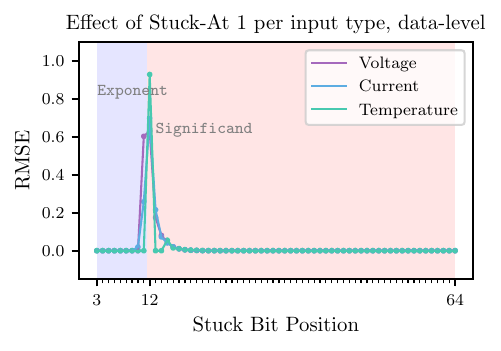}
     \caption{}
     \label{fig:data1}
     \end{subfigure}%
     \caption{RMSE between original and corrupted data for stuck-at faults injected into bits 3-64 of each input type, which are in turn a series of \texttt{Float64}. \ref{fig:data0}: stuck-at 0; \ref{fig:data1}: stuck-at 1.}
     \label{fig:data}
\end{figure}

Figures \ref{fig:data0} and \ref{fig:data1} show the RSME deviations not on the level of the model's predictions, but on the level of deviations between the original data and the corrupted data. These results mirror those from the predictions above and show that large deviations in the predictions are tied to large deviations in the data. Furthermore, it appears that even though the effect of stuck-at-0 on temperature is the largest on a data level, on a prediction level it is the smallest (see Figure \ref{fig:rmse0}). This suggest that an internal (and opaque) weighting of inputs by an AI model can make it difficult to predict behaviour from only looking at the input data.

\begin{figure}[h]
    \centering
    \includegraphics{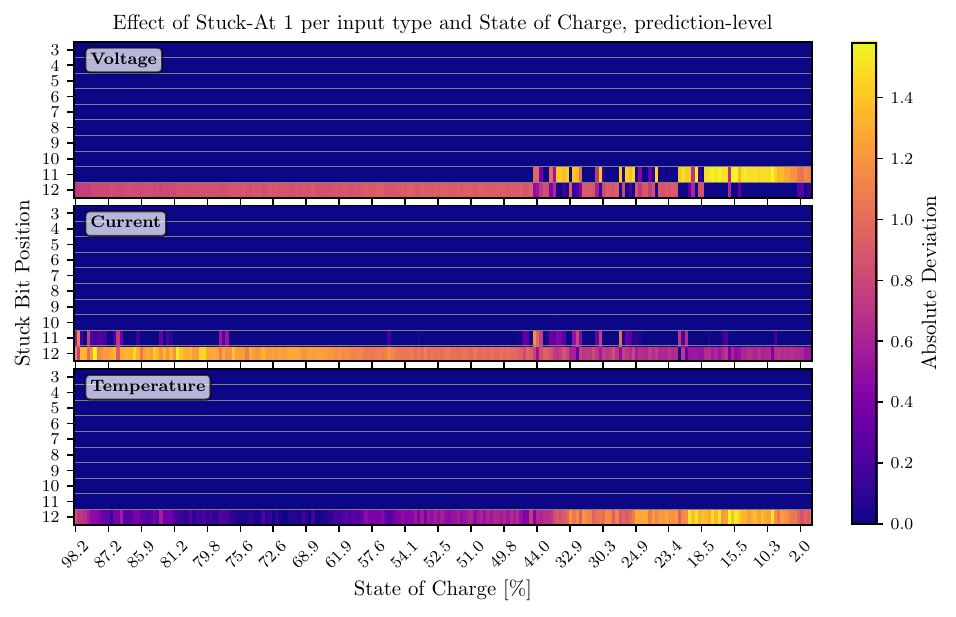}
    \caption{Absolute Deviation between SOC predictions from original and corrupted data throughout the discharge cycle. Stuck-at 1 faults injected into bits 3-12 of each input type, which are in turn a series of \texttt{Float64}.}
    \label{fig:abs_error_1}
\end{figure}

Figure \ref{fig:abs_error_1} illustrates the Absolute Deviation for each prediction of SOC throughout the discharge cycle for stuck-at-1 faults. This visualisation enables the analysis of the model's behaviour at various stages of the discharge cycle. For clarity, the figure displays only the effects on exponent bits (bits 3 to 12).

As previously mentioned, bits 3 to 10 already have the value 1 at the start and are unaffected by the injected stuck-at-1 faults; thus, no sensitivity is observed for these bits as a function of SOC percentage. However, significant influences from the Voltage input begin at bit 11 and below 44\% charge. For Current, the influence of bit 11 is less pronounced but more spread out throughout the discharge cycle. For Temperature, only bit 12 has an effect.

Additionally, the largest zone of impact appears at different locations for each input type. For Voltage, this occurs at bit 11 below approximately 44\% charge. For Current, it occurs at bit 12 above roughly 72.6\% charge, and for Temperature, at bit 12 below 32.9\% charge.

The behaviour observed in Figure \ref{fig:abs_error_1} aligns with the results presented earlier in Figure \ref{fig:data1}, where the RMSE value increases only when a high exponent bit (11 or 12) is stuck-at-1. The mechanism through which this affects the model's predictions is challenging to determine without a thorough understanding of its inner workings. However, as previously stated, the different input types (Voltage, Current, Temperature) may be subjected to various internal weighting. Both Figures \ref{fig:rmse} and \ref{fig:abs_error_1} demonstrate that significant deviations from the original predictions can occur if the exponent bits of the model's inputs are altered.

To understand the varying impact of stuck-at faults on the Absolute Deviation at different locations for each input type in Figure \ref{fig:abs_error_1}, one needs to examine the changes in \texttt{Float64} values throughout the discharge cycle. Although all model inputs are normalised within the range $[0;1]$, as stated earlier, they can still differ in how much of this interval they utilise. 
For \texttt{Float64} in the range $(1;0)$, it is true that bits 3-10 are all 1, while bits 11 and 12 are $10_2$ for values in $(1;0.5]$ and $01_2$ for values in $(0.5;0)$. During our tests, we observed that Temperature remains relatively stable throughout the discharge cycle with a value close to 1. In contrast, Voltage and Current fall from values close to 1 to values below 0.5 during the discharge cycle. Thus, the sensitivity to stuck-at-1 faults changes depending on the data value, explaining the patterns seen in Figure \ref{fig:abs_error_1}: Temperature is only sensitive in bit 12, while Voltage and Current are sensitive even in bit 11 if their value falls in the interval $(0.5,0)$.

\paragraph{Summary}
The experiments underline that stuck-at faults can cause significant deviations from the AI component's original predictions, potentially resulting in undesired behaviour, such as markedly inaccurate SOC predictions. These findings highlight the critical importance of detecting errors like stuck-at faults in the input data using the monitor system proposed in Section \ref{ai_monitor}.

\section{Conclusions and future work}

This work addresses the challenges of ensuring the safety of AI-based systems, focusing on SOC estimation. The main contribution, besides an introduction for integrating AI in the V-Model for safety assessments, is the identification of the architectural element of the safety cage as a good candidate for demarcation and interface between traditional, AI-independent measures aligned with ISO 26262 and AI-dependent measures requiring alignment with emerging standards like ISO/PAS 8800.
Our experiments underline how sensitive an AI SOC prediction model can be to common faults such as \textit{stuck-at}.

% Our experimental setup is designed to explore various aspects of the AI SOC estimator in the presence of faults. Specifically, the following work is considered:

% \begin{itemize}
%     %\item Analyse the results obtained using additional fault models, including the various failure modes for sensors listed in ISO 26262-5.
%     \item Analyse the results obtained by incorporating additional fault models, taking into account the various failure modes, \ref{AIC1}, \ref{AIC2}, \ref{AIC3}, identified as relevant to this type of AI-based sensor.
%     \item Investigate the workload impact by varying the time intervals for the injected faults and applying different charge/discharge cycles.
%     \item Evaluate the safety monitor proposed in Section \ref{ai_monitor}.
%     \item Evaluate the quality and relevance of using training data with faults already applied in order to improve the robustness of the AI-driven SOC estimations.
% \end{itemize}

Future work will enhance the multi-concern assessment framework \cite{templates_2023} by integrating new AI findings and refining methods to address data quality, in-service monitoring, and fallback strategies for AI-based systems.
Furthermore, we will analyse the effects of additional failure modes from \ref{AIC1}, \ref{AIC2}, \ref{AIC3} on AI SOC prediction, and investigate the quality and relevance of using training data with faults already applied in order to improve the robustness of the AI-driven SOC estimations. Finally, it will evaluate the effectiveness of the safety monitor itself.

\section*{Acknowledgments} 
We acknowledge the support of the Swedish Knowledge Foundation via the industrial doctoral school RELIANT, grant nr: 20220130. This research was carried out within the SUNRISE project and is funded by the European Union’s Horizon Europe Research and Innovation Actions under grant agreement No. 101069573. However, views and opinions expressed are those of the author(s) only and do not necessarily reflect those of the European Union or the European Union’s Horizon Europe Research and Innovation Actions.

%For bibliography, please use the attached bst file or format as follows:

%\begin{thebibliography}{99}
\bibliographystyle{evs.bst}   % the inclu∂ed bst file
%Zotero EVS38 in the RISE of mean lean paper machine
\bibliography{./ref/EVS38.bib}

\begin{thebibliography}{10}
\providecommand{\url}[1]{#1}
\csname url@samestyle\endcsname
\providecommand{\newblock}{\relax}
\providecommand{\bibinfo}[2]{#2}
\providecommand{\BIBentrySTDinterwordspacing}{\spaceskip=0pt\relax}
\providecommand{\BIBentryALTinterwordstretchfactor}{4}
\providecommand{\BIBentryALTinterwordspacing}{\spaceskip=\fontdimen2\font plus
\BIBentryALTinterwordstretchfactor\fontdimen3\font minus \fontdimen4\font\relax}
\providecommand{\BIBforeignlanguage}[2]{{%
\expandafter\ifx\csname l@#1\endcsname\relax
\typeout{** WARNING: IEEEtran.bst: No hyphenation pattern has been}%
\typeout{** loaded for the language `#1'. Using the pattern for}%
\typeout{** the default language instead.}%
\else
\language=\csname l@#1\endcsname
\fi
#2}}
\providecommand{\BIBdecl}{\relax}
\BIBdecl

\bibitem{iso26262}
{International Organization for Standardization (ISO)}, ``{ISO 26262:2018 Road Vehicles – Functional Safety}.''

\bibitem{iso21448}
------, ``{ISO 21448:2022 Road vehicles — Safety of the intended functionality}.''

\bibitem{ISOPAS8800}
\BIBentryALTinterwordspacing
------. {ISO/PAS 8800 Road vehicles — Safety and artificial intelligence}. ISO. [Online]. Available: \url{https://www.iso.org/standard/83303.html}
\BIBentrySTDinterwordspacing

\bibitem{natm}
ECE/TRANS/WP.29/2021/61, ``({{GRVA}}) {{New Assessment}}/{{Test Method}} for {{Automated Driving}} ({{NATM}}) - {{Master Document}} | {{UNECE}},'' World Forum for Harmonization of Vehicle Regulations.

\bibitem{henriksson_out--distribution_2023}
J.~Henriksson, S.~Ursing, M.~Erdogan, F.~Warg, A.~Thorsén, J.~Jaxing, O.~Orsmark, and M.~O. Toftås, ``Out-of-distribution detection as support for autonomous driving safety lifecycle,'' in \emph{Requirements Engineering: Foundation for Software Quality}, A.~Ferrari and B.~Penzenstadler, Eds.\hskip 1em plus 0.5em minus 0.4em\relax Springer Nature Switzerland, pp. 233--242.

\bibitem{ali_analyzing_2020}
\BIBentryALTinterwordspacing
N.~Ali, M.~Hussain, and J.-E. Hong, ``Analyzing {Safety} of {Collaborative} {Cyber}-{Physical} {Systems} {Considering} {Variability},'' \emph{IEEE Access}, vol.~8, pp. 162\,701--162\,713, 2020. [Online]. Available: \url{https://ieeexplore.ieee.org/document/9186018/}
\BIBentrySTDinterwordspacing

\bibitem{salay_safety_2019}
\BIBentryALTinterwordspacing
R.~Salay, M.~Angus, and K.~Czarnecki, ``A {Safety} {Analysis} {Method} for {Perceptual} {Components} in {Automated} {Driving},'' in \emph{2019 {IEEE} 30th {International} {Symposium} on {Software} {Reliability} {Engineering} ({ISSRE})}.\hskip 1em plus 0.5em minus 0.4em\relax Berlin, Germany: IEEE, Oct. 2019, pp. 24--34. [Online]. Available: \url{https://ieeexplore.ieee.org/document/8987559/}
\BIBentrySTDinterwordspacing

\bibitem{qi_hierarchical_2022}
\BIBentryALTinterwordspacing
Y.~Qi, P.~R. Conmy, W.~Huang, X.~Zhao, and X.~Huang, ``A {Hierarchical} {HAZOP}-{Like} {Safety} {Analysis} for {Learning}-{Enabled} {Systems},'' 2022, version Number: 1. [Online]. Available: \url{https://arxiv.org/abs/2206.10216}
\BIBentrySTDinterwordspacing

\bibitem{adee_systematic_2021}
\BIBentryALTinterwordspacing
A.~Adee, R.~Gansch, and P.~Liggesmeyer, ``Systematic {Modeling} {Approach} for {Environmental} {Perception} {Limitations} in {Automated} {Driving},'' in \emph{2021 17th {European} {Dependable} {Computing} {Conference} ({EDCC})}.\hskip 1em plus 0.5em minus 0.4em\relax Munich, Germany: IEEE, Sep. 2021, pp. 103--110. [Online]. Available: \url{https://ieeexplore.ieee.org/document/9603704/}
\BIBentrySTDinterwordspacing

\bibitem{adee_discovery_2021}
\BIBentryALTinterwordspacing
A.~Adee, R.~Gansch, P.~Liggesmeyer, C.~Glaeser, and F.~Drews, ``Discovery of {Perception} {Performance} {Limiting} {Triggering} {Conditions} in {Automated} {Driving},'' in \emph{2021 5th {International} {Conference} on {System} {Reliability} and {Safety} ({ICSRS})}.\hskip 1em plus 0.5em minus 0.4em\relax Palermo, Italy: IEEE, Nov. 2021, pp. 248--257. [Online]. Available: \url{https://ieeexplore.ieee.org/document/9660641/}
\BIBentrySTDinterwordspacing

\bibitem{berk_assessing_2020}
\BIBentryALTinterwordspacing
M.~Berk, O.~Schubert, H.-M. Kroll, B.~Buschardt, and D.~Straub, ``Assessing the {Safety} of {Environment} {Perception} in {Automated} {Driving} {Vehicles},'' \emph{SAE International Journal of Transportation Safety}, vol.~8, no.~1, pp. 49--74, 2020, publisher: SAE International. [Online]. Available: \url{https://www.jstor.org/stable/27034112}
\BIBentrySTDinterwordspacing

\bibitem{demirci_review_2024}
O.~Demirci, ``Review of battery state estimation methods for electric vehicles - part i: {SOC} estimation.''

\bibitem{neto2022}
A.~V.~S. Neto, J.~B. Camargo, J.~R. Almeida, and P.~S. Cugnasca, ``Safety {{Assurance}} of {{Artificial Intelligence-Based Systems}}: {{A Systematic Literature Review}} on the {{State}} of the {{Art}} and {{Guidelines}} for {{Future Work}},'' \emph{IEEE Access}, vol.~10, pp. 130\,733--130\,770, 2022.

\bibitem{avizienis_basic_2004}
A.~Avizienis, J.-C. Laprie, B.~Randell, and C.~Landwehr, ``Basic concepts and taxonomy of dependable and secure computing,'' \emph{{IEEE} Transactions on Dependable and Secure Computing}, vol.~1, no.~1, pp. 11--33, 2004.

\bibitem{sangchoolie2018}
B.~Sangchoolie, P.~Folkesson, and J.~Vinter, ``A study of the interplay between safety and security using model-implemented fault injection,'' in \emph{2018 14th European Dependable Computing Conference (EDCC)}.\hskip 1em plus 0.5em minus 0.4em\relax IEEE, 2018, pp. 41--48.

\bibitem{qutub_hardware_2022}
S.~Qutub, F.~Geissler, Y.~Peng, R.~Gräfe, M.~Paulitsch, G.~Hinz, and A.~Knoll, ``Hardware faults that matter: Understanding and estimating the safety impact of hardware faults on object detection {DNNs},'' in \emph{Lecture Notes in Computer Science (including subseries Lecture Notes in Artificial Intelligence and Lecture Notes in Bioinformatics)}, vol. 13414 {LNCS}, 2022, pp. 298--318.

\bibitem{beyer_online_2023}
M.~Beyer, J.~Borrmann, A.~Guntoro, and H.~Blume, ``Online quantization adaptation for fault-tolerant neural network inference,'' in \emph{Lecture Notes in Computer Science (including subseries Lecture Notes in Artificial Intelligence and Lecture Notes in Bioinformatics)}, vol. 14181 {LNCS}, 2023, pp. 243--256.

\bibitem{ruiz_zero-space_2024}
J.~Ruiz, D.~De~Andres, L.~Saiz-Adalid, and J.~Gracia-Moran, ``Zero-space in-weight and in-bias protection for floating-point-based {CNNs},'' in \emph{Proceedings - 2024 19th European Dependable Computing Conference, {EDCC} 2024}, 2024, pp. 89--96.

\bibitem{10.1145/3462203.3475878}
\BIBentryALTinterwordspacing
K.~L. Wong, M.~Bosello, R.~Tse, C.~Falcomer, C.~Rossi, and G.~Pau, ``{Li-Ion Batteries State-of-Charge Estimation Using Deep LSTM at Various Battery Specifications and Discharge Cycles},'' in \emph{Proceedings of the Conference on Information Technology for Social Good}, ser. GoodIT '21.\hskip 1em plus 0.5em minus 0.4em\relax New York, NY, USA: Association for Computing Machinery, 2021, p. 85–90. [Online]. Available: \url{https://doi.org/10.1145/3462203.3475878}
\BIBentrySTDinterwordspacing

\bibitem{Mendeley}
\BIBentryALTinterwordspacing
P.~Kollmeyer, C.~Vidal, M.~Naguib, and M.~Skells. (2020) {LG 18650HG2 Li-ion Battery Data and Example Deep Neural Network xEV SOC Estimator Script}. {Version} 3. [Online]. Available: \url{https://data.mendeley.com/datasets/cp3473x7xv/3}
\BIBentrySTDinterwordspacing

\bibitem{templates_2023}
\BIBentryALTinterwordspacing
M.~Skoglund, F.~Warg, A.~Thorsén, and M.~Bergman, ``Enhancing safety assessment of automated driving systems with key enabling technology assessment templates,'' \emph{Vehicles}, vol.~5, no.~4, pp. 1818--1843, 2023. [Online]. Available: \url{https://www.mdpi.com/2624-8921/5/4/98}
\BIBentrySTDinterwordspacing

\end{thebibliography}
\small

%\end{thebibliography} 
%\newpage
\bigskip
\section*{Presenter Biography}
\begin{minipage}[b]{21mm}
\includegraphics[width=20mm]{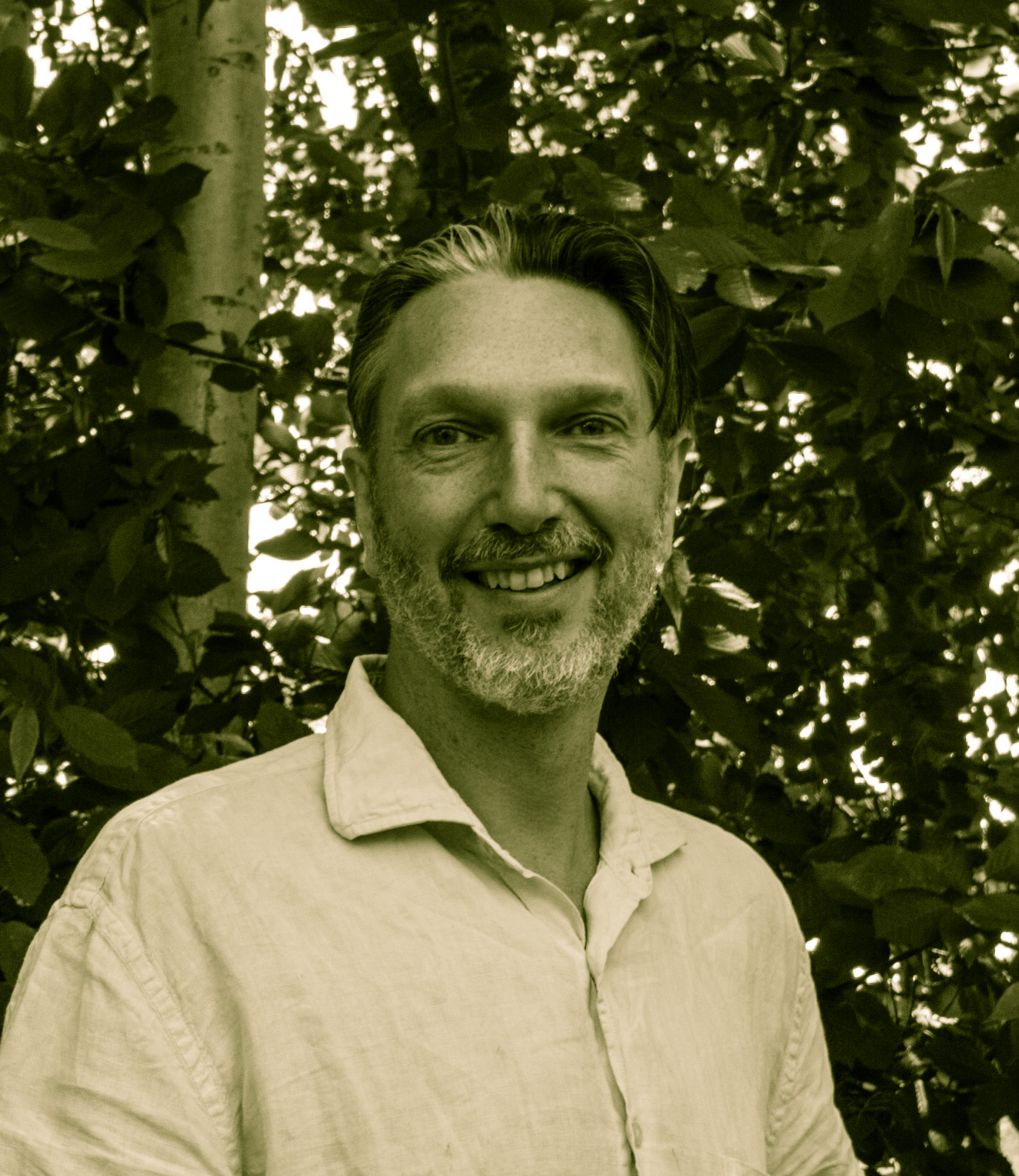}
\end{minipage}
\hfill
\begin{minipage}[b]{140mm}
\small
Martin Skoglund works within the Dependable Transport Systems unit at RISE Research Institutes of Sweden in Borås, focusing on safety assurance for connected and automated systems through advancing methods for evaluating safety and security-informed safety. His expertise encompasses functional safety, cybersecurity, safety of the intended functionality, and artificial intelligence, with research aimed at improving the efficiency and effectiveness of safety assessments to ensure safe and correct operation under all relevant conditions.
\end{minipage}
\\\\

\end{document}